\documentclass[letterpaper]{aa520}

\usepackage{graphicx}
\usepackage{txfonts}
\begin{document}

\title{Odin observations of the Galactic centre in the 118-GHz
band\thanks{Based on observations with Odin, a Swedish-led satellite
  project funded jointly by the Swedish National Space Board (SNSB),
  the Canadian Space Agency (CSA), the National Technology Agency of
  Finland (Tekes) and Centre National d'Etude Spatiale (CNES). The
  Swedish Space Corporation was the industrial prime contractor
  and is also responsible for the satellite operation.}}

\subtitle{Upper limit to the O$_{2}$\ abundance}

\author{Aa.\,Sandqvist\inst{1}
  \and B.\,Larsson\inst{1}
  \and \AA .\,Hjalmarson\inst{2}
  \and P.\,Bergman\inst{2}
  \and P.\,Bernath\inst{3}
  \and U.\,Frisk\inst{4}
  \and M.\,Olberg\inst{2}
  \and L.\,Pagani\inst{5}
\and L.M.\, Ziurys\inst{6} 
}

\institute{Stockholm Observatory, AlbaNova University Center, SE-106
91 Stockholm, Sweden\\ \email{aage@astro.su.se}
  \and Onsala Space Observatory, SE-439 92 Onsala, Sweden
  \and Department of Chemistry, University of York, Heslington, York,
  YO10 5DD, UK
  \and Swedish Space Corporation, P. O. Box 4207, SE-171 04 Solna,
  Sweden
  \and LERMA \&\  UMR8112 du CNRS, Observatoire de Paris, 61,
  FR-75014 Paris, France
  \and Arizona Radio Observatory, Steward Observatory, University of
  Arizona, Tucson, AZ 85721, USA   
}

\offprints{\\ Aage Sandqvist, \email{aage@astro.su.se}}

\date{Received 22 January 2008 / Accepted 28 February 2008}

\abstract {} {The Odin satellite has been used to search for the 118.75-GHz
  line of molecular oxygen (O$_{2}$) in the Galactic centre.} {Odin
  observations were performed towards the Sgr\ts$\rm {A}^{*}$\ circumnuclear
  disk (CND), and the Sgr\ts A\ +20 \,{km\,s$^{-1}$}\ and +50
  \,{km\,s$^{-1}$}\ molecular clouds 
  using the position-switching mode. Supplementary ground-based
  observations were carried out in the 2-mm band using the ARO Kitt
  Peak 12-m telescope to examine suspected SiC features.} {A strong
  emission line was found at 118.27 GHz, attributable to the $J=13-12$
  HC$_{3}$N\ line. Upper limits are presented for the 118.75-GHz O$_{2}$\
  $(1_{1}-1_{0})$ ground transition line and for the 118.11-GHz
  $^{3}\Pi_{2}, J=3-2$\ ground state SiC line at the Galactic
  centre. Upper limits are also presented for the 487-GHz O$_{2}$\ line in
  the Sgr\ts A\  +50 \,{km\,s$^{-1}$}\  cloud and for the 157-GHz,
  $J=4-3$, SiC line in the 
  Sgr\ts A\   +20 and +50 \,{km\,s$^{-1}$}\ clouds, as well as the
  CND. The CH$_{3}$OH 
  line complex at 157.2 - 157.3 GHz has been detected in the +20 and
  +50 \,{km\,s$^{-1}$}\  clouds but not towards Sgr\ts$\rm
  {A}^{*}$/CND.} {A $3\sigma$\ 
  upper limit for the fractional abundance ratio of [O$_{2}$]/[H$_2$] is
  found to be $X$(O$_{2}$) $\le 1.2 \times 10^{-7}$ towards the Sgr\ts A\
  molecular belt region.}  

{   

  \keywords{Galaxy: center -- ISM: individual objects: Sgr\ts A\  -- ISM:
  molecules  -- ISM: abundances -- ISM: clouds} 
}

\maketitle

\authorrunning{Aa. Sandqvist et al.}

\section{Introduction}

The Odin submillimetre/millimetre wave spectroscopy astronomy and
aeronomy satellite has been in operation since 2001. The satellite has
been described in detail by Frisk et al. (\cite{fri03})  and the
receiver calibration by Olberg et al. (\cite{olb03}). Its prime
astronomical priorities were to search the interstellar medium for
water vapour (H$_{2}$O ) at a frequency of 556.93600 GHz and
molecular oxygen (O$_{2}$ ) at a frequency of 118.75034 GHz, both
frequencies not observable from the ground. The H$_{2}$O\
observations have been abundantly successful, a summary of which has
been presented by Hjalmarson et al. (\cite{hja03}, \cite{hja07}). Emission and
absorption in the 557-GHz H$_{2}^{16}$O\  line have been detected in the
Galactic centre towards the Sgr\ts$\rm {A}^{*}$\ circumnuclear disk (CND), and
the Sgr\ts A\ +20 \,{km\,s$^{-1}$}\ and +50 \,{km\,s$^{-1}$}\
molecular clouds (Sandqvist et 
al. \cite{san03}, \cite{san06}). Good reviews on the Galactic centre
Region have been presented by Morris \& Serabyn (\cite{mor96}) and
Mezger et al. (\cite{mez96}). Upper limits have been achieved for the O$_{2}$\
abundances in a number of regions, our first upper limits being of the
order of $\leq 10^{-7}$ (Pagani et al. \cite{pag03}). However, for one
source, namely the $\rho$ Oph\, A molecular cloud, a very weak O$_{2}$\
line has actually been detected (Larsson et al. \cite{lar07}).

In this paper, we report the results of Odin observations using the
on-board 118-GHz radiometer made towards the Sgr\ts A\ts complex\  in the
Galactic centre.

\section{Observations}

The observations were performed during 26 March - 9 April 2003 and 17
- 26 February 2004 and were part of a larger Galactic centre project
The receiver used was the 118-GHz cryo-cooled HEMT receiver with a
single-sideband (SSB) temperature of  $\approx730$ K in 2003 and
$\approx840$ K in 2004. The backend spectrometer for the 118-GHz
receiver consisted of an acousto-optical spectrometer (AOS) with a
total bandwidth of 1.05 GHz and a channel resolution of 1 MHz. At the
118-GHz frequency, the angular resolution of the Odin beam is about
$10'$ and the main beam efficiency is 0.91 (Frisk et al. \cite{fri03}).  

Three positions were observed, separated by a few
arcminutes. The total-power position-switching observing mode was used
with a duty cycle of 120 s, the empty reference OFF-position being 
at $\alpha$(B1950.0)=17$^{\rm h}40^{\rm m}$26$^{\rm s}$.8, $\delta$(B1950.0)=
$-28\degr35'$04\arcsec. The 1950.0 coordinates of the three source
positions are given in Table 1, and Fig. 1 shows these positions
superimposed upon a continuum/molecular distribution map of 
the Sgr\ts A\ts complex. Most of the features in the Sgr\ts A\ts
complex\  seen in this 
figure are within the 118-GHz $10'$ -beam at all three positions. Due
to different problems with the AOS in the 2003 period only 12 orbits (hours) of
observation produced useful data (eight towards Sgr\ts$\rm {A}^{*}$\
CND, three towards 
the +20 \,{km\,s$^{-1}$}\ cloud and one towards the +50
\,{km\,s$^{-1}$}\ cloud). During the 
2004 period a total of 77 orbits produced useful data (divided equally
amongst the three above-mentioned positions).   

\begin{table}
\caption{Observed Positions in the Galactic centre Sgr\ts A\ region}
\begin{flushleft}
\begin{tabular}{lll}
\hline\noalign{\smallskip}
\hline\noalign{\smallskip}
      &  $\alpha$(B1950.0)      &   $\delta$(B1950.0) \\
\hline\noalign{\smallskip}
Sgr\ts$\rm {A}^{*}$\  circumnuclear disk & 17$^{\rm h}$42$^{\rm m}$29$^{\rm
s}$.3  &  $-28\degr$$59'$18\arcsec \\
Sgr\ts A\  +20 \,{km\,s$^{-1}$}\  cloud  & 17$^{\rm h}$42$^{\rm
  m}$29$^{\rm s}$.3 
&  $-29\degr$$02'$18\arcsec \\
Sgr\ts A\  +50 \,{km\,s$^{-1}$}\  cloud  & 17$^{\rm h}$42$^{\rm
  m}$41$^{\rm s}$.0 
&  $-28\degr$$58'$00\arcsec \\
\noalign{\smallskip}\hline\end{tabular}
\end{flushleft}
\end{table}

\begin{figure}
%  \resizebox{\hsize}{!}{\rotatebox{0}{\includegraphics{odinfig1.2ps}}}
  \caption{{\it This 350-K figure can be obtained by downloading the
      ``odinfig1.gif'' file. It is an adaption of Fig. 7 in Sandqvist
      (1989, A\&A 223, 293)}
 The three observed positions in the Galactic centre Sgr\ts A\
  region are marked by crosses, the black cross indicating also
  Sgr\ts$\rm {A}^{*}$. The Sgr\ts A\ts complex\  consists of the
  Sgr\ts$\rm {A}^{*}$, 
  Sgr\ts A\ts West, Sgr\ts A\ts East\ and Sgr\ts A\ts halo\ radio
  continuum sources (20-cm 
  continuum radiograph), the circumnuclear disk (CND, HCN - thin
  contours), and the molecular belt (2-mm H$_{2}$CO\ - thick
  contours). Isovelocity contours are indicated with dashed contours.
  The straight diagonal line indicates the orientation of the Galactic
  plane. The coordinates are epoch 1950.0. (adapted from Sandqvist
  \cite{san89})}   
  \label{1}
\end{figure}

The 118-GHz receiver on board Odin is not properly phase-locked but
slowly drifts in frequency. However, during 40\,\% of Odin's observing
time, the satellite is looking at the Earth's atmosphere. This
allows us to monitor the stability of the receiver and to
correct for any frequency drift using the telluric oxygen
line. Larsson et al. (\cite{lar03}, \cite{lar07}) have given a
detailed description of this method of frequency calibration.

In February/March 2007, Odin observed the Sgr\ts A\  +50
\,{km\,s$^{-1}$}\  cloud 
during 150 orbits with one of the front end submillimetre receivers
tuned to 486.79638 GHz, the rest frequency of the $(3_3 - 1_2)$ O$_{2}$\
line, using the position-switching method. The back end spectrometer
was an autocorrelator (AC2) with a total bandwidth of 700 MHz and a
channel resolution of 1.2 MHz (0.62 \,{km\,s$^{-1}$}). The SSB system
temperature was 
about 3000 K. The angular resolution of the Odin beam at this
frequency is $2\farcm39$ and the main beam efficiency is 0.89.  

Supplementary ground-based observations in the 157-GHz band were
performed with 12-m millimetre wave telescope of the
Arizona Radio Observatory at Kitt Peak in March 2004 to check a
suspicious feature seen in the 2003 Odin 119-GHz profile near the
frequency expected for a possible SiC feature (see Fig. 1.1 of
Hjalmarson et al. \cite{hja04}). The 2-mm receiver (133-180 GHz), with
two orthogonal front ends, was tuned to 157.4941010 GHz, the expected
rest frequency of the $^{3}\Pi_{2}, J = 4 - 3$ SiC line (Pickett et
al. \cite{pic98}). It had a nominal SSB receiver temperature of about
125 K. The total system temperature varied between 600 and 1500 K,
depending on the elevation of the source during the observations. The
backend spectrometer was the Millimeter Autocorrelator (MAC) operated in
parallel mode with useable bandwidth of 600 MHz and an effective
channel resolution of 0.74 \,{km\,s$^{-1}$}. Again the position-switching
observation mode was used. The beamwidth of the 12-m telescope at 157
GHz is 40\arcsec.

\section{Results}

The 2004 observations at the three positions were co-added after the careful
frequency calibrations outlined in the previous section. The resulting
profile is shown in Fig. 2. The intensity scale has not been corrected for
the main beam efficiency ($\approx 0.9$) and a weak second-order
polynomial baseline has been subtracted. The AOS has some
low-level system instability which is variable in amplitude and
frequency over the band. To obtain a signal-free OFF position in the
Galactic centre region, a large switching angle (0\fdg6) had to be
used resulting in a switching period of 120 s. Due to this, some of
the system instability may not have been completely removed,
causing a non-uniform rms noise level across the AOS band.

The strong line present in Fig. 2 we propose is identified as the $J=13-12$
transition of the HC$_{3}$N\ (cyanoacetylene) molecule which has a rest
frequency of 118.2707322 GHz. This line was also very clear in the
2003 data, which has been published in a preliminary report on Odin
observations by Hjalmarson et al. (\cite{hja04}). In that profile
there appeared also to be a weak broader feature at the lower
frequency side of the HC$_{3}$N\  line which was suspected to possibly be
the $^{3}\Pi_{2}, J=3-2$\  ground state transition of the SiC molecule
with a rest frequency of 118.1122437/93 GHz. At the high frequency end
of the observed band of that profile there was also a hint of emission
from the O$_{2}$\ molecule whose $(1_{1}-1_{0})$ transition has a rest
frequency of 118.7503430 GHz. All quoted rest frequencies are taken
from the on-line JPL Molecular Spectroscopy Catalogue at
http://spec.jpl.nasa.gov/ (Pickett et al. \cite{pic98}). In the much
more sensitive profile of the 2004 data presented here in Fig. 2,
there are no obvious indications of these SiC and O$_{2}$\ lines (whose
expected frequency intervals are marked by the individual velocity
scales) and we deem that uncertainties in the baseline caused these
suspected features in the 2003 data profile.

\begin{figure}
  \resizebox{\hsize}{!}{\rotatebox{90}{\includegraphics{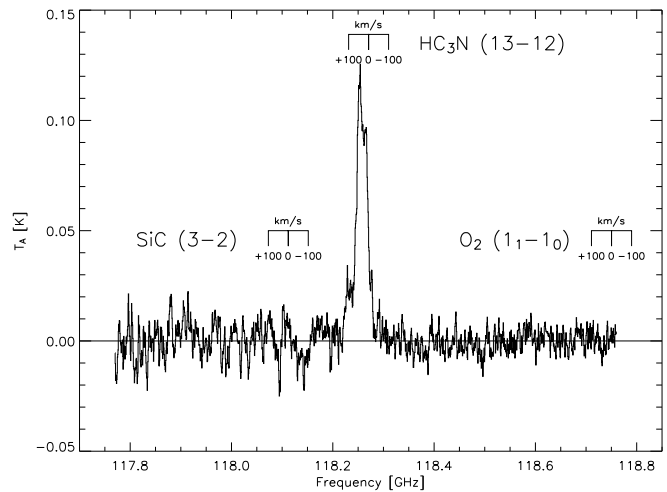}}}
  \caption{Odin 118-GHz observations of the Sgr\ts A\ts complex\  region in the
  Galactic centre - total integrated profile, channel resolution: 1.6
  \,{km\,s$^{-1}$}. Small velocity scales are shown above each
  expected position of 
  line features 
 }
\label{2}
\end{figure}

\subsection{HC$_{3}$N}

The clarity and sharpness of the strong HC$_{3}$N\ feature in Fig. 2 shows
that the frequency calibration of the 118-GHz receiver can be
considered successful. The feature is also strong enough to permit the
limited mapping done in connection with the submillimetre
observations. The individual HC$_{3}$N\ spectra observed towards the
Sgr\ts$\rm {A}^{*}$\ CND, the +20 and +50 \,{km\,s$^{-1}$}\ clouds,
together with the total 
integrated profile, are shown in Fig. 3. 

The Odin $10'$-beamwidth covers practically the whole region shown in
Fig. 1 and the map spacing is smaller than
thirdbeam-spacing. It is clear from both the velocity halfwidths (which
are of the order of 50 \,{km\,s$^{-1}$}) and the velocities of the
peak temperatures that 
the HC$_{3}$N\  profiles show a behaviour consistent with their dominant
sources being the +20 and +50 \,{km\,s$^{-1}$}\ clouds in the
molecular belt (cf 
the H$_{2}$CO\ profiles in Sandqvist \cite{san89}). These are relatively
warm ($\approx100$ K), high-density ($\approx10^{4.5}$) giant molecular
clouds with moderate velocity dispersion ($\approx25$ \,{km\,s$^{-1}$}) which,
however, have only few  signs of star formation.

\begin{figure}
  \resizebox{\hsize}{!}{\rotatebox{90}{\includegraphics{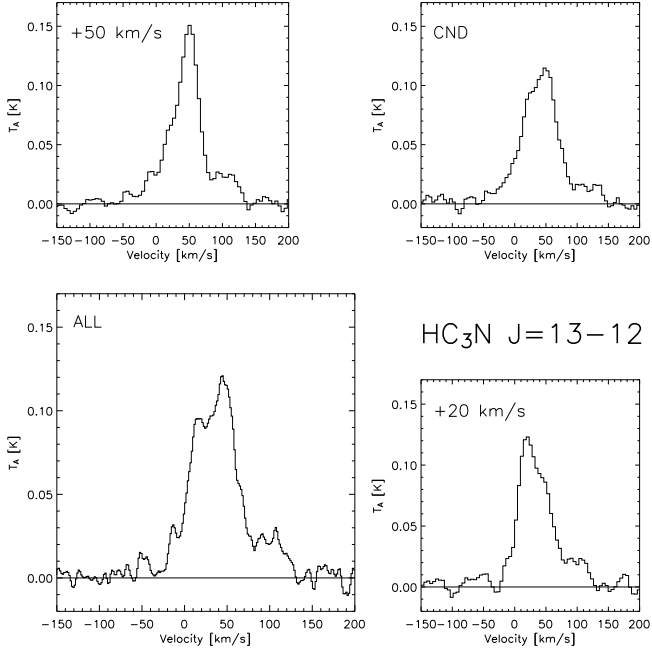}}} 
  \caption{HC$_{3}$N\ ``map'' showing the three profiles obtained towards
  the +50 \,{km\,s$^{-1}$}\ cloud (upper left), Sgr\ts$\rm {A}^{*}$\ CND (upper
  right) and the +20\,{km\,s$^{-1}$}\ cloud (lower right) - channel
  resolution is 
  4.7 \,{km\,s$^{-1}$}. The total integrated profile with a channel
  resolution of 
  1.6 \,{km\,s$^{-1}$}\ is shown in the lower left.}
\label{3}
\end{figure}

\subsection{O$_{2}$}

No O$_{2}$\ line was detected down to a  $T_{\rm mb}$ rms limit of 4.8
mK, a value 
determined for the frequency range of 118.400 - 118.752
GHz. Unfortunately, the slow frequency drift of the Odin 119-GHz
receiver had moved the O$_{2}$\ frequency to the very edge of the
observing band in 2004 and thus only the positive O$_{2}$\ velocity range (and
a small amount of the negative O$_{2}$\ velocity range) was
observable. However, molecular emission from the Sgr\ts A\ts complex\ dominates
in the positive velocity range (see eg. Fig. 3). So, if any O$_{2}$\ were
present in the Sgr\ts A\ts complex\ it would have been included in this
profile. At this frequency, the 10 arcmin Odin beam includes all three
Sgr\ts A\ts complex\ components (see Fig. 1).

The 487-GHz O$_{2}$\ observations, on the other hand, have an angular
resolution of only $2\farcm4$ and are thus limited to the component at which
Odin is pointing. The profile observed towards the Sgr\ts A\  +50
\,{km\,s$^{-1}$}\ 
cloud is presented in Fig. 4 and it is quite featureless. This profile
has been boxcar-smoothed and no O$_{2}$\ line was detected down to a
$T_{\rm mb}$ rms limit of 9.0 mK.

\begin{figure}
  \resizebox{\hsize}{!}{\rotatebox{270}{\includegraphics{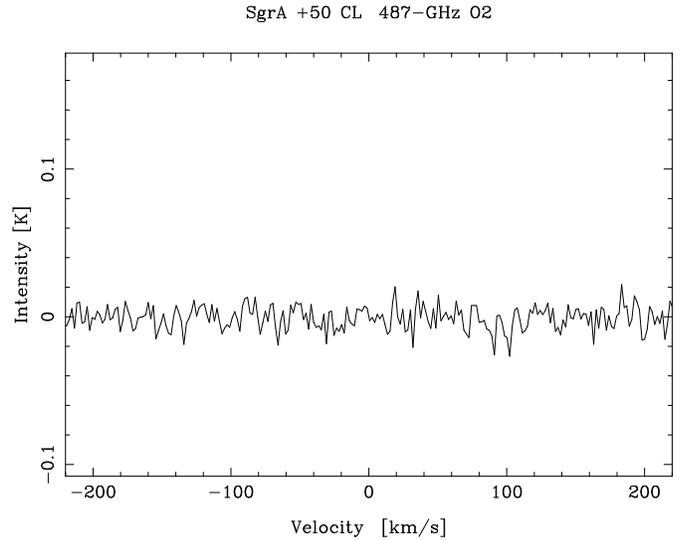}}}
  \caption{Odin 487-GHz O$_{2}$\ observations of the Sgr\ts A\  +50
    \,{km\,s$^{-1}$}\ 
  cloud. Channel resolution: 1.8 \,{km\,s$^{-1}$}. 
 }
\label{4}
\end{figure}

\subsection{SiC and CH$_{3}$OH}

It is quite clear from Fig. 2 that there is no obvious detection of
118-GHz ground-state emission from SiC in the Sgr\ts A\ts complex. Due to the
non-uniformity of the rms noise across the Odin 118-GHz band (see
above), calculating a  $T_{\rm mb}$ rms limit for the ground state SiC
line was limited to the frequency range 117.8 - 118.2 GHz, yielding a
value of 8.7 mK. 
 
The attempt to observe the higher SiC transition with the ARO 12-m
mm-wave telescope was motivated by the noisy 2003 Odin observations as
mentioned in Sect. 2 and was carried out before we had access to the
2004 Odin observations. Although no detection of a SiC line was made,
we nonetheless present the results of these ARO observations
here. The observations made with the two orthogonal front ends were
averaged together. These observations did serendipitously result in
the detection of the strong CH$_{3}$OH line complex at 157.2 - 157.3
GHz in the +20- and +50-\,{km\,s$^{-1}$}\  clouds but {\it not} towards the
CND. The lack of any CH$_{3}$OH emission towards the CND position is not really
surprising since the 40\arcsec\  beam of the ARO 12-m telescope
would be mostly looking at the cavity inside the CND. The baselines
including a dominant standing wave were removed by assuming a
featureless profile towards the CND cavity, - based on the lack of any
CH$_{3}$OH feature towards the CND as mentioned above -
super-smoothing it by Hanning it 19 times, and then subtracting the
result from all three profiles. The results are shown in Fig. 5. The
$T_{\rm mb}$ rms limits in the three profiles are 20, 11 and 19 mK in
the +50 \,{km\,s$^{-1}$}\ cloud, CND and +20 \,{km\,s$^{-1}$}\ cloud,
respectively, assuming a 
main beam efficiency of 0.7.   

\begin{figure}
  \resizebox{\hsize}{!}{\rotatebox{270}{\includegraphics{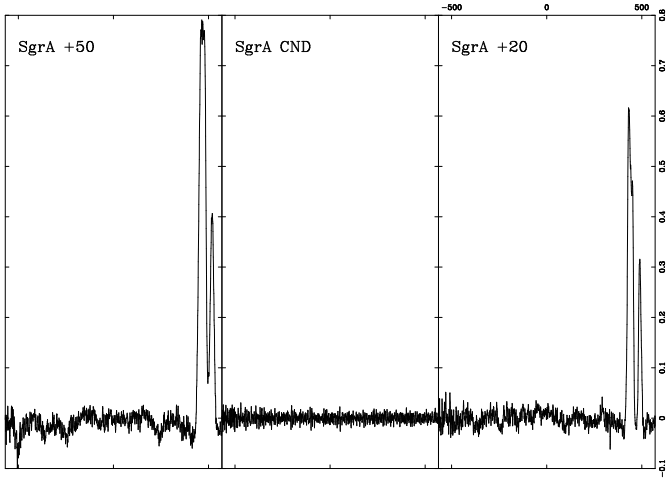}}}
  \caption{ARO 12-m telescope 157-GHz observations of the Sgr\ts A\ts complex\
  region in the Galactic centre - towards the +50 \,{km\,s$^{-1}$}\
  cloud, the CND 
  and the +20 \,{km\,s$^{-1}$}\ cloud. The velocity x-axis ranges from
  $-550$ to +550 \,{km\,s$^{-1}$}\ 
  assuming the SiC rest frequency as the reference and the intensity
  y-axis ranges from $-0.1$ to +0.8 K. Channel resolution: 0.74
  \,{km\,s$^{-1}$}. The strong lines near +500 \,{km\,s$^{-1}$}\
  originate from the CH$_{3}$OH 
  line complex at 157.2 - 157.3 GHz. 
 }
\label{5}
\end{figure}

\section{Discussion}
 
\subsection{HC$_{3}$N}

The HC$_{3}$N\ molecule has many transitions which can be observed from the
ground and it is considered to be an excellent density probe with
optically thin lines (Vanden Bout et al. \cite{van83}). Walmsley et
al. (\cite{wal86}) have made measurements of nine of these transitions
towards selected regions in the Sgr\ts A\  molecular belt, ranging in frequency
from 9 to 226 GHz. From a statistical equilibrium model they derived a
density of $10^{4}$ cm$^{-3}$ for the +20 and +50 \,{km\,s$^{-1}$}\ clouds
assuming a temperature of 80 K. The fit could be slightly improved
non-uniquely by assuming that 20\%\ of the mass is in clumps with a
density of $10^{5}$ cm$^{-3}$ with the remaining 80\%\  at a density
of $5 \times 10^{3}$ cm$^{-3}$. 

Two of their transitions, ($12-11$) and ($15-14$), stradle our
($13-12$) transition but they were observed with the Texas 4.9-m Millimetre
Wave Observatory (MWO) telescope which has five times better resolution
($\approx 2$\ arcmin) than Odin. The Odin profiles towards the +20 and
+50 \,{km\,s$^{-1}$}\ clouds have intensity maxima at radial
velocities of +20 and +50 
\,{km\,s$^{-1}$}, respectively. The integrated line intensity of the Odin
$J=13-12$ HC$_{3}$N\ total profile (ALL in Fig. 3) was found to be
about 8.8 K \,{km\,s$^{-1}$}\ assuming a main beam efficiency of
0.91. The elongated 
shape and angular dimensions of the molecular belt are well
established from different molecular observations (e.g. Fig. 1). We
will here approximate the size with a {\it spherical}, homogeneous cloud of
diameter of 6.4 arcmin or 16 pc, assuming a distance of 8.5 kpc to the
Galactic centre. This corresponds to an Odin beam dilution of
0.25. Correcting for this beam dilution we match the modelled
$J=13-12$ intensity of about 35 K \,{km\,s$^{-1}$}\ of Walmsley et
al. (\cite{wal86}). The  
Walmsley et al. best-fit model we adopt here is their single-component
model having the parameters: gas temperature $T = 80$ K, uniform gas density
$n_{\rm H2}=10^{4}$ cm$^{-3}$, and a fractional HC$_{3}$N\ abundance per
velocity gradient of $10^{-9}$ (\,{km\,s$^{-1}$}\ pc$^{-1})^{-1}$. We find then
a molecular hydrogen column density through the centre of the model
cloud $N$(H$_{2}$) of $5 \times 10^{23}$ cm$^{-2}$, and since the
whole model cloud is encompassed by our beam we observe a total mass
of $10^{6}$ M$_{\sun}$. Lis \&\ Carlstrom (\cite{lis94}) find
$N$(H$_{2}$) values of $4.1 \times 10^{23}$ cm$^{-2}$ for the +20
\,{km\,s$^{-1}$}\ 
cloud and $2.4 \times 10^{23}$ cm$^{-2}$ for the +50 \,{km\,s$^{-1}$}\
cloud from their observations of the 800-$\mu$m continuum emission
and they deduce masses of $5 \times 10^{5}$ M$_{\sun}$ and $4 \times
10^{5}$ M$_{\sun}$, respectively, for these two components of the
molecular belt. Finally, by using a velocity gradient of 2 \,{km\,s$^{-1}$}\
pc$^{-1}$ (cf. the isovelocity contours of Fig. 1) we estimate the
HC$_{3}$N\ fractional abundance to be $2 \times 10^{-9}$. Hence the
central HC$_{3}$N\ column density for the molecular belt model source is
$N$(HC$_{3}$N)$= 10^{15}$ cm$^{-2}$.

\subsection{O$_{2}$}

In order to make an estimate of the upper limit of the O$_{2}$\ abundance,
implied by our non-detections of the 119- and 487-GHz lines, we use as
the starting point the HC$_{3}$N\ modelling results of Walmsley et
al. (\cite{wal86}) and our results for the HC$_{3}$N\ $J=13-12$ line as
presented in the previous section. We then adjust the O$_{2}$\ abundance
in our model cloud until we obtain a predicted maximum intensity for
the 119-GHz O$_{2}$\ line, as observed with the Odin 10\arcmin\ beam,
which is compatible with the rms noise level of our observation. Our
4.8 mK rms noise value of the non-detected 119-GHz O$_{2}$\ observation
then yields a $3\sigma$\  upper limit for the O$_{2}$\ fractional
abundance of $X$(O$_{2}$) $\le 1.2 \times 10^{-7}$. The $3\sigma$\  upper
limit for the central O$_{2}$\ column density in the model
cloud is $N$(O$_{2}$) $\le 6 \times 10^{16}$ cm$^{-2}$. This model also
results in a predicted 487-GHz O$_{2}$\ line main beam temperature of
$T_{\rm mb} \approx 1$ mK, well below the rms noise of 9 mK in our
487-GHz observation.     

The first Odin O$_{2}$\ non-detection results improved the earlier SWAS
upper limits in star forming regions by a factor of a few and in cold dark
clouds by factors of  between 20 and 40 (Pagani et al. \cite{pag03};
Goldsmith et al. \cite{gol00}). The Odin results gave O$_{2}$\ abundance
upper limits of $X$(O$_{2}$) between $5 \times 10^{-8}$ and $8 \times
10^{-7}$ for different Galactic sources. In the Small Magellanic Cloud,
Wilson et al. (\cite{wil05}) found an upper limit of the O$_{2}$\
abundance ratio of $1.3 \times 10^{-6}$. The only source in
which O$_{2}$\ has actually been detected, through its 119-GHz transition,
is $\rho$ Oph A, where Odin found an O$_{2}$\ abundance ratio of $X$(O$_{2}$)
$= 5 \times 10^{-8}$ (Larsson et al. \cite{lar07}).  

The Sgr\ts A\ts complex\ contains a variety of conditions potentially
favourable for relatively high O$_{2}$\ abundance. One such region is the
existence of the shock front interaction between the expansion of the
Sgr\ts A\ts East\ shell into the molecular belt and the centrally peaked
thermal X-ray emission from the interior of the Sgr\ts A\ts East\ shell
(Maeda et al. \cite{mae02}). The O$_{2}$\ could be a tracer of the X-rays
in the warm envelope of the molecular belt facing Sgr\ts A\ts East\ (see e.g.
Goldsmith et al. \cite{gol02} for a discussion on shock-produced O$_{2}$\
or St\"auber et al. \cite{sta05} for a discussion on X-ray-induced chemical
models). On the other hand, the UV radiation field emitted by the central
cluster of hot massive stars (the IRS16 stars) (Allen et
al. \cite{all90}) strikes the surface of the molecular belt with the
result that the surface of the +50\,{km\,s$^{-1}$}\ cloud which faces
Sgr\ts$\rm {A}^{*}$\ 
and IRS16 is bright in [CII] line emission - as if this part of the
cloud is directly photo-dissociated by the UV radiation from the
Centre itself (Genzel et al. \cite{gen90}). This would have a negative
effect on the production of O$_{2}$\ in this part of the molecular belt's
envelope. But all this would only be valid for the relatively thin
inner envelope of the molecular belt, much beyond the resolution of
the Odin beam. 

Consider now the whole molecular belt. Unlike many other giant
molecular clouds in the Galaxy, the molecular belt cloud components do 
not contain any significant embedded massive star formation regions
which could induce increased O$_{2}$\ abundance by heating surrounding
dust complexes. In fact, it has been shown in a number of observations
that, again differing from many other Galactic regions, the Sgr\ts
A\ts complex\ 
molecular clouds have considerably lower dust temperatures than gas
temperatures (e.g. G\"usten et al. \cite{gus85}). So, instead of 80 K,
which is the gas temperature that we use in our abundance
determination above, the actual dust temperature is more like 30 K
(Lis \&\ Carlstrom \cite{lis94}).  

Hollenbach et al. (\cite{hol08}) have studied the oxygen chemistry in
clouds irradiated by an external FUV flux. This model includes
photodissociation, standard gas phase chemistry, freezeout of species
on grain surfaces, some limited grain surface chemistry and some
desorption mechanisms. Hollenbach et al. have kindly applied their
model to our case. In their standard case for a cloud with density of
$10^{4}$ cm$^{-3}$ and an external FUV field which is 100 times
greater than the local ISRF, they obtain at the {\it surface} a gas
temperature of 120 K and a dust temperature of 31 K, but at the place
where the H$_{2}$O\ and the O$_{2}$\ abundances peak they get a gas
temperature of 22 K and a dust temperature of 15 K. Increasing the
external radiation field until they get a dust temperature of 30 K at
the O$_{2}$\ peak, they find that the column density of O$_{2}$\ in the cloud
has a value of $N$(O$_{2}$) = $2 \times 10^{16}$ cm$^{-2}$. This is a
factor of three lower than our upper limit. So, it is not unreasonable
that we were unable to detect the O$_{2}$\ line, assuming the validity of
these models.

\subsection{SiC}

It was not until 1987 that the SiC molecule was spectroscopically
identified in the laboratory (Bernath et al. \cite{ber88}). The SiC
molecule has not previously been observed in the interstellar 
medium. Only in the circumstellar envelope of the carbon star IRC +10216 and
the evolved carbon star CIT-6 has SiC been detected, and then not at
the ground-level frequency (Cernicharo et al. \cite{cer89}). In a
manner similar to the SiO molecule, it may be that strong shocks
disrupting dust grains are required to release SiC into the gaseous
interstellar medium. Martin-Pintado et al. (\cite{mar97}) have
observed that SiO emission in the Galactic centre regions seems to
delineate closely the nonthermal Radio Arc and also there is a region
of enhanced SiO emission close to Sgr\ts$\rm {A}^{*}$. There is
nonetheless also 
SiO emission coming from the +20 and +50 \,{km\,s$^{-1}$}\ clouds. It is not
unreasonable to suspect that SiC may coexist in the SiO regions. 

The upper limits implied by the non-detection of SiC at 118 GHz and
157 GHz were determined from the spectroscopic data of the Cologne
Database for Molecular Spectroscopy (M\"uller et al. \cite{mul05}). The
line intensities are based on an {\it ab initio} dipole moment of 1.7 D
from an early calculation (Cernicharo et al. \cite{cer89}). Note that
both the Cologne and JPL databases are essentially identical for
SiC. A more recent {\it ab initio} calculation gives a very similar
value of 1.65 D for the equilibrium dipole moment (Pramanik \& Das
\cite{pra07}), so the Einstein A values given by the Cologne database were used
($A_{3-2}=1.268 \times 10^{-5}$ s$^{-1}$ and $A_{4-3}=4.385 \times
10^{-5}$ s$^{-1}$).

Assuming an excitation temperature of 80 K and the corresponding
partition function of 280 gives upper limit column densities of
$16 \times 10^{13}$ molecules cm$^{-2}$ for the $^{3}\Pi_{2}, J=3-2$,
118 GHz line (25\% beam dilution) and $4 \times 10^{13}$ molecules
cm$^{-2}$ for $J=4-3$ at 157 GHz (no beam dilution). These values are
based on limiting RMS noise levels of 8.7 mK and 20 mK over 100
\,{km\,s$^{-1}$\ for Odin and ARO, respectively, and taking the two unresolved
$\Lambda$-doublets of equal intensity into account. These upper limits
compare favourably with the observed column density of $6 \times 10^{13}$
molecules cm$^{-2}$ in IRC+10216 (Cernicharo et al. \cite{cer89}). A
3$\sigma$ upper limit for the fractional abundance ratio of
[SiC]/[H$_{2}$] in the Sgr\ts A\ molecular belt is then estimated to be
$X$(SiC) $\le 2 \times 10^{-10}$, using the H$_{2}$ column density
derived in Sect. 4.1.   

From our upper limit to the SiC column density of $4 \times 10^{13}$
cm$^{-2}$ and the SiO column density range $(7-30) \times 10 ^{13}$
cm$^{-2}$ determined by Martin-Pintado er al. (\cite{mar97}) we arrive
at an upper limit to the SiC/SiO abundance ratio $(0.13 - 0.57)$ which
is just similar to the standard C/O elemental abundance ratio (0.5)
listed by Grevesse et al. (\cite{gre96}). Hence deeper SiC searches
would be desirable.

\section{Conclusions}

Molecular oxygen remains as elusive as ever, the Galactic centre
appears to be no exception. Despite long integrations using the Odin
satellite, the 118.75-GHz ground state O$_{2}$\ line has not been
detected towards the Sgr\ts A\ts complex, nor has the 487-GHz O$_{2}$\
line. We find a 
$3\sigma$\  upper limit for the O$_{2}$\ column density in the Sgr\ts A\
molecular belt of $N$(O$_{2}$) $\le 6 \times 10^{16}$ cm$^{-2}$, which is
compatible with the results of new theoretical oxygen chemistry
models. The $3\sigma$\  upper limit for the fractional abundance ratio of
[O$_{2}$]/[H$_{2}$] is $X$(O$_{2}$) $\le 1.2 \times 10^{-7}$.

\begin{acknowledgements}

We should like to express our gratitude to D. Hollenbach et al. for having
applied their model to the Sgr\ts A\ts complex\ molecular belt and for
having made 
available results to us before publication. We also thank the referee
for constructive comments. The Kitt Peak 12-m millimetre wave
telescope is operated by the Arizona Radio Observatory (ARO), Steward
Observatory, University of Arizona.  

\end{acknowledgements}

\end{document}